\documentclass[twocolumn,aps,pra,showpacs,superscriptaddress,
amsfonts,amssymb,amsmath]{revtex4}
\usepackage{mathrsfs}
\usepackage{amsmath}
\usepackage{graphicx}
\usepackage{float}

\begin{document}
\draft
\title{Excitation Spectra and Hard-core Thermodynamics of Bosonic Atoms in Double-well Optical Lattices}

\author{Bao Xu}
\author{Hong-Ye Wu}
\affiliation{The Department of Physics Science and Technology, Baotou Normal College, Baotou 014030, China}

\author{Yi-Cai Zhang}
\author{Han-Ting Wang}
\author{Wu-Ming Liu}
\affiliation{Beijing National Laboratory for Condensed Matter Physics, Institute of Physics, 
  Chinese Academy of Sciences, Beijing 100190, China}

\begin{abstract}

A generalized coupled representation is proposed for bosonic atoms 
in double-well optical lattices. The excitation spectra and thermodynamic properties of these systems are investigated in the coupled representation.
The excitation processes with filling factor one 
can be described by simultaneously creating
doubly occupied and empty double wells. 
Then it is demonstrated that hard-core statistics must be taken into account to properly describe the equilibrium properties at finite temperatures. 
Finally, the cases with other filling factors are also briefly discussed. 

\end{abstract}


\maketitle

 %%%%%%%%%%%%%%%%%%%%%%%%%%%%%%%%%%%%%%%%%%%%%%%%%%%%%

\section{introduction}

 Optical lattices are flexible to manipulate spatial dimensions,
 topological structures, well depth and periodic length.
 The influence of periodic potential on particles can be 
 systematically simulated via atom gases in optical lattice.
 These optical lattice systems  therefore  provide new 
 opportunities to explore open questions in strongly correlated physics 
 due to limited adjustability of crystal lattices.    
 Transitions from Mott-insulator to superfluid phases were experimentally realized in 2002 
 by continuously changing well depths of the optical lattice \cite{greiner}.
 Since then, intriguing experimental and theoretical advances have been made  
 in optical-lattice systems to mimicking conventional 
 strongly correlated physics \cite{lewenstein,bloch2008a,yukalov09}.
 
 Multi-well optical lattices exhibit more versatile physical 
 properties than ordinary optical lattices \cite{santos2004}. 
 The simplest configuration among those 
 are double-well lattices. 
 Interesting dynamic behaviours of cold atoms in one double well, 
 such as quantum interference between two fragments of BEC,
 Josephson tunnelling through the central barrier, 
 self-trapping and co-tunnelling of atom pairs  in strongly  correlated regime, \cite{folling} 
 have been theoretically and experimentally investigated.
 Double-well lattices has been experimentally realized 
 by J. Sebby-Strabley et al. \cite{sebby06} and subsequently
 attracted considerable attention \cite{danshita,hepb,jiangsj,stojanovi,yukalov}.
 The tunnelling amplitude through central barrier, 
 interaction strength between atoms, and depth imbalance (tilt) of double wells
 can be manipulated by changing the intensity or relative phase of laser 
 standing waves that engineer the optical lattice. 
 The contact interaction via effective scattering length $a_{s}$  
 can also be changed by ramping applied magnetic fields
 relative to the Feshbach-resonance field $B_{c}$ \cite{greiner,petrov,wouters}.
 The experimental control of superexchange interactions in double-well ladders was
 reported in Refs. \cite{lmduan2003,trotzky,yachen2011}. 
 In addition, successful controls of atomic pairs in double-well lattices provide 
 new promising candidates for quantum computation,
 quantum information processing and quantum communication \cite{sebby07,anderlini,bloch2008}.

 Different from ordinary optical lattices, single mode approximation is not suitable for double-well lattices.
 Two or even more modes are required to describe the dynamic and equilibrium properties of bosonic atoms 
 in double-well lattices.
 It is reasonable to expect certain advantages to study such systems in coupled representations, 
 as shown in spin-dimer systems.
 In those systems, each spin operator can be expressed in spin coupled representation. 
 This method has been proved to be powerful to understand experimental results 
 in spin-dimer systems \cite{sachdev,kumar,nikuni,ruegg03,xu}.
 To the best of our knowledge, there is still lack of generalized coupled representations 
 for cold atoms in double-well lattices. 
 The first part of this paper attempts to address the issue.

 Secondly, this coupled representation is applied to hard-core bosons 
 in double-well lattices.
 When repulsions between bosonic atoms are enormously strong,
 no more than one atom can stay in one well simultaneously.
 The repulsion plays a similar role as 
 the Pauli exclusion principle for fermions. 
 Such bosonic atoms are referred to as hard-core bosons.
 Although the ground state phase diagram of these atoms in double-well
 lattices have been investigated for various cases,
 the effects of finite temperatures have not been explored extensively.
 As will be demonstrated, hard-core statistics  
 must be taken into account to properly describe  
 the properties at finite temperatures,
 such as temperature dependence of heat capacity and level 
 occupation numbers.

 This paper is organized as follows: 
 In section II, a generalized coupled
 representation for bosonic atoms in double-well lattices is proposed, 
 and subsequently applied to the hard-core case.
 Section III presents the model Hamiltonian discussed in this paper
 and derives its effective expression in the coupled representation.
 Then the hard-core statistics for bosonic atoms in double-wells lattices 
 is derived. Subsequently the self-consistent saddle-point equations
 for the effective Hamiltonian are obtained.  
 In Section IV, the excitation spectra and hard-core thermodynamic 
 properties are discussed. 
 The paper is summarized in Section V.
 
%%%%%%%%%%%%%%%%%%%%%%%%%%%%%%%%%%%%%%%%%%%%%%%%%%%%%%%%%%%%%%%%%%%

\section{coupled representations for bosonic atoms in double-well lattices}

\subsection{Generalized Form of Coupled Representation}

 The system of cold atoms confined in a double-well lattice mathematically
 resembles a spin-dimer system \cite{matsuda}, of which the ground state
 can be constructed from spin-dimer singlets. 
 The latter system has been extensively studied
 because the Bose-Einstein Condensation of triplons may be 
 detected in such magnetic systems \cite{giamarchi}.
 In this section, the generalized formalism of the coupled representation for 
 bosonic atoms in double-well lattices is presented. 
 
 The uncoupled bases of double-well bosonic atoms 
 can be expressed in terms of particle occupations 
 $|n_{\rm L},n_{\rm R}\rangle$, 
 where ${\rm L}$ and ${\rm R}$ represent 
 the left and right sides of a double well, respectively.  
 Assume the left (right) side of the double well
 can accommodate no more than $2j_1$ ($2j_2$) bosons, with
 $j_1$ ($j_2$) being integers or half odd integers.
 Then the uncoupled bases of the double well can be
 denoted by $|n_{\rm L},n_{\rm R}\rangle$ with $0\leq n_{\rm L}\leq2j_1$ and $0\leq n_{\rm R}\leq2j_2$.
 It is noticed that the bases $|n_{\rm L},n_{\rm R}\rangle$
 correspond to that of the uncoupled representation $|j_1,j_2,m_1,m_2\rangle$ 
 for two spins operators, $\vec{j}_1$ and $\vec{j}_2$, 
 with $j^{z}$ components of each spin equal to 
 $m_1 = n_{\rm L} - j_1$ and $m_2 = n_{\rm R} - j_2$, respectively.
 It is known that the uncoupled representation of two
 spins connects to their coupled representation through the Clebsch-Gordan coefficients.
 In the same way, we can build the relationship between 
 the coupled and uncoupled representations for bosonic atoms in double-well lattices.

 Firstly, we can express the creation and annihilation operators 
 of the bosonic atoms in the left (right)
 side of a double well $b^{\dag}_{\rm L}$,$b_{\rm L}$ ($b^{\dag}_{\rm R}$,$b_{\rm R}$)
 using the Hubbard X-operators expressions  
 $|n'_{\rm L},n'_{\rm R}\rangle\langle n_{\rm L},n_{\rm R}|
 \equiv {\rm X}^{(n'_{\rm L},n'_{\rm R})(n_{\rm L},n_{\rm R})}$ 
 in the uncoupled representation as \cite{hubbard}
 \begin{eqnarray}
 &&b^{\dag}_{\rm L} = \sum_{n_{\rm L},n_{\rm R}}\sqrt{n_{\rm L}+1}|n_{\rm L}+1,n_{\rm R} \rangle\langle n_{\rm L},n_{\rm R}|,\nonumber\\
 &&b^{\dag}_{\rm R} = \sum_{n_{\rm L},n_{\rm R}}\sqrt{n_{\rm R}+1}|n_{\rm L},n_{\rm R}+1 \rangle\langle n_{\rm L},n_{\rm R}|,\nonumber\\
 &&b_{\rm L} = ( b^{\dag}_{\rm L} )^{\dag},~~~~ b_{\rm R} = ( b^{\dag}_{\rm R} )^{\dag}.\label{blr}
 \end{eqnarray}

 Secondly, the coupled bases of the bosonic atoms in double wells are denoted by
 $|n:\alpha\rangle$. 
 Both $n$ and $\alpha$ are necessary and also sufficient to span the local Hilbert space 
 of atomic occupations.
 The total number of atoms in a double well $n=n_{\rm L}+n_{\rm R}$
 corresponds to the $z$-component of the total spins $m=m_1+m_2$.
 The parameter $\alpha=j_1+j_2+J$, stemming from the total spin  
 $|j_1-j_2|\leq J\leq j_1+j_2$, reflects the symmetry of the double-well bosonic basis vectors.  
 It is straightforward to explicitly write down the double-well bosonic atom bases $|n:\alpha\rangle$
 by reference to the representation of two spins
 $|j_1,j_2,J,m\rangle$.  
 We introduce {\it Y-operator}, which is similar to the X-operator defined by Hubbard,
 using the double-well coupled bases mentioned above, 
 ${\rm Y}^{(n',\alpha')(n,\alpha)} \equiv |n':\alpha' \rangle\langle n:\alpha|$.
 Define ${\rm Y}^{(n',\alpha')(n,\alpha)} = a^{\dag}_{n',\alpha'}a_{n,\alpha}$,
 where $a^{\dag}_{n,\alpha}$ ($a_{n,\alpha}$) represents creating (annihilating) 
 a bosonic atom in the double-well coupled bases $|n:\alpha\rangle$.

 Finally, the creation and annihilation operators in Eqn. (\ref{blr}) can be
 rewritten in the coupled representation as
 \begin{eqnarray}
  && b^{\dag}_{\rm L} = 
  \sum_{n_1,n_2,j,l}  C^{j,n_1-j_1+1+n_2-j_2}_{n_1-j_1+1,n_2-j_2}
   C^{l,n_1-j_1+n_2-j_2}_{n_1-j_1,n_2-j_2}\nonumber\\
  && ~~~ \times ~ \sqrt{n_1+1}~ a^{\dag}_{n_1+n_2+1:j+j_1+j_2} a_{n_1+n_2:l+j_1+j_2}; \label{bs1}\\
 %%%%%%%%%%%%%%%%%%%%
 && b^{\dag}_{\rm R} = \sum_{n_1,n_2,j,l}C^{j,n_1-j_1+n_2-j_2+1}_{n_1-j_1,n_2-j_2+1}
    C^{l,n_1-j_1+n_2-j_2}_{n_1-j_1,n_2-j_2}\nonumber\\
 && ~~~ \times ~ \sqrt{n_2+1} ~ a^{\dag}_{n_1+n_2+1:j+j_1+j_2} a_{n_1+n_2:l+j_1+j_2}; \label{bs2}\\
 %%%%%%%%%%%%%%%%%%%% 
 && b_{\rm L} = ( b^{\dag}_{\rm L} )^{\dag},~~~~ b_{\rm R} = ( b^{\dag}_{\rm R} )^{\dag}.
 \end{eqnarray}
 where 
 $0\leq n_1\leq 2j_1$, $0\leq n_2\leq 2j_2$, 
 $|j_1-j_2|\leq j,l\leq j_1+j_2$, 
 and $C^{j,m}_{m_{1},m_{2}}$ are the Clebsch-Gordan coefficients.
 It is ready to check that above expressions satisfy the
 ordinary bosonic commutation relations, as long as the constrained condition
 \begin{eqnarray}
 \sum_{n,\alpha}a^{\dag}_{n,\alpha}a_{n,\alpha}=1    \label{hardcore}
 \end{eqnarray}
 is satisfied.
 This constrained condition implies that the double well can only be in
 one of the orthogonal bases that span the local Hilbert space. 
 % read as follows
 % \begin{widetext}
 % \begin{eqnarray}
 % C^{j,m}_{m_{1},m_{2}}&=&
 % \delta_{m,m_{1}+m_{2}}\sqrt{\frac{(j_{1}+j_{2}-j)!(j+j_{1}-j_{2})!(j+j_{2}-j_{1})!(2j+1)}
 % {(j+j_{1}+j_{2}+1)!}}\nonumber\\
 % &&\times\sum_{k}\frac{(-1)^{k}\sqrt{(j_{1}+m_{1})!
 % (j_{1}-m_{1})!(j_{2}+m_{2})!(j_{2}-m_{2})!(j+m)!(j-m)!}}
 % {k!(j_{1}+j_{2}-j-k)!(j_{1}-m_{1}-k)!(j_{2}+m_{2}-k)!(j-j_{2}+m_{1}+k)!(j-j_{1}-m_{2}+k)!},\nonumber\\
 % \end{eqnarray}
 % \end{widetext}
 %with ${\rm max}\{0,-j+j_{2}-m_{1},-j+j_{1}+m_{2}\}\leq k\leq {\rm min}\{j_{1}+j_{2}-j,j_{1}-m_{1},j_{2}+m_{2}\}$.

%%%%%%%%%%%%%%%%%%%%%%%%%%%%%%%%%%%%%%%%%%%%%%
%%%-- hard-core bosonic atoms--------%%%%%%%%%
%%%%%%%%%%%%%%%%%%%%%%%%%%%%%%%%%%%%%%%%%%%%%%%

\subsection{Coupled Representation in the Hard-core Limit}

For clarity, we begin with the simplest case for atoms in one double well, 
considering tunnelling amplitude through the central barrier ($t$), 
intra-well ($U$)  and inter-well ($U_{1}$) repulsive 
interactions.  
For the system with one particle in the double well,
the model Hamiltonian, involving the tunnelling term ($t$) only , 
can be readily diagonalized.
Eigenvalues are $-t$ and $t$,
and the corresponding coupled bases are symmetric 
$|s\rangle=\frac{1}{\sqrt{2}}(  |1_{\rm L}0_{\rm R}\rangle + |0_{\rm L}1_{\rm R}\rangle  )$,
and antisymmetric 
$|a\rangle=\frac{1}{\sqrt{2}}(  |1_{\rm L}0_{\rm R}\rangle - |0_{\rm L}1_{\rm R}\rangle  )$,
respectively.
The symmetric basis $|s\rangle$ corresponds to the lower energy level, since $t>0$.
The above scheme to obtain coupled bases $|s\rangle$ and $|a\rangle$ 
is similar to the method to construct 
molecular orbits of H$^{+}_{2}$ from atomic orbits of hydrogen H \cite{sarma}.
The symmetric (antisymmetric) basis here corresponds to the bonding (antibonding) orbit.  

If there are two particles in the double well with $t$, $U$  and $U_{1}$ finite, 
the uncoupled bases are 
$|1_{\rm L} 1_{\rm R}\rangle$, $|0_{\rm L} 2_{\rm R}\rangle$ and $|2_{\rm L} 0_{\rm R}\rangle$.
Diagonalizing the model Hamiltonian gives three eigenvalues 
$E_{1}=U$, $E_{2,3}=  \frac{U+U_{1}}{2}  \mp \sqrt{ \left(\frac{U-U_{1}}{2}\right)^{2}  + 2t^{2} }   $, 
and the corresponding coupled bases read as 

\begin{eqnarray}
\phi_{2}^{(1)} &=& \frac{1}{ \sqrt{2} } \left( 
|2_{\rm L} 0_{\rm R}\rangle - |0_{\rm L} 2_{\rm R}\rangle
\right),\nonumber\\
%%%%%%%
\phi_{2}^{(2)} &=&  |2_{\rm L} 0_{\rm R}\rangle + |0_{\rm L} 2_{\rm R}\rangle \nonumber\\
&&  +~\left[ \frac{U-U_{1}}{2t}  + \sqrt{ \left(\frac{U-U_{1}}{2t}\right)^{2}  + 2 }~
\right]                    																
| 1_{\rm L} 1_{\rm R} \rangle                    \nonumber\\
%%%%%%%
\phi_{2}^{(3)} &=& |2_{\rm L} 0_{\rm R}\rangle + |0_{\rm L} 2_{\rm R}\rangle             \nonumber\\
&&	+~\left[ \frac{U-U_{1}}{2t}  - \sqrt{ \left(\frac{U-U_{1}}{2t}\right)^{2}  + 2 }~
\right]																		
|1_{\rm L} 1_{\rm R}\rangle ;	       					\nonumber
\end{eqnarray}  
where the normalization constants in $\phi_{2}^{(2)}$ and $\phi_{2}^{(3)}$ have been omitted for brevity.

In the specific case when $U\gg t$ and $U\gg U_{1}$, 
energy of $\phi_{2}^{(2)}$ equals $-2t^{2}/(U-U_{1})+U_{1}$,  
which is much less than those of the other two bases, namely, 
$U$ of $\phi_{2}^{(1)}$, and
$2t^{2}/(U-U_{1})+U$ of $\phi_{2}^{(3)}$, respectively.
It is clear that the first term $-2t^{2}/(U-U_{1})$   
reflects contributions of second-order perturbation processes from 
$|1_{\rm L} 1_{\rm R}\rangle$ to $|0_{\rm L} 2_{\rm R}\rangle$ and $|2_{\rm L} 0_{\rm R}\rangle$. 
It is also noticed $|d\rangle\equiv|1_{\rm L} 1_{\rm R}\rangle$ 
is the dominant component of $\phi_{2}^{(2)}$   
since $(U-U1)/t\to\infty$ in the case when $U\gg t$ and $U\gg U_{1}$. 
$|1_{\rm L} 1_{\rm R}\rangle$  is therefore mostly occupied at vanishing temperatures.

 We now can conclude that the most relevant bases for bosonic atoms in the double-well potential are
 $|s\rangle$,$|a\rangle$, and $|d\rangle\equiv|1_{\rm L} 1_{\rm R}\rangle$. 
 In order to satisfy the completeness of state space, another basis 
 $|e\rangle\equiv|0_{\rm L}0_{\rm R}\rangle$, {\it i.e.}, no atoms in the double well, has to be included.  
 The generalized coupled representation in Eqn.s (\ref{bs1}) and (\ref{bs2}) 
 now reduces to the following form
 
 \begin{eqnarray}
 &&b^{\dagger}_{\rm L,R} = \frac{1}{\sqrt{2}}[d^{\dagger}(s\mp a) + (s^{\dagger} \pm  a^{\dagger})e],\label{dwbo}
 \end{eqnarray}
 where 
 $d^{\dagger}|\phi\rangle = |1_{\rm L}1_{\rm R}\rangle$ 
 represents the doubly occupied double-well (dDW) basis 
 with $|\phi\rangle$ the Fock vacuum;
 $e^{\dagger}|\phi\rangle = |0_{\rm L}0_{\rm R}\rangle$ represents 
 the empty double-well (eDW) basis;
 $s^{\dagger}|\phi\rangle = \frac{1}{\sqrt{2}}(|1_{\rm L}0_{\rm R}\rangle + |0_{\rm L}1_{\rm R}\rangle )$ 
 denotes singly occupied symmetric double-well (sDW) basis and  
 $a^{\dagger}|\phi\rangle = \frac{1}{\sqrt{2}}(|1_{\rm L}0_{\rm R}\rangle - |0_{\rm L}1_{\rm R}\rangle )$ 
 expresses singly occupied antisymmetric double-well (aDW) basis.
 
 When the coupling between double-wells vanishes, 
 zero-point energy per double well are 
 $-2t^{2}/(U-U_{1})+U_{1}$ (dDW), $0$ (eDW), $-t_1$ (sDW) and
 $t_1$ (aDW), respectively.
 The constraint condition in Eqn. (\ref{hardcore}) is now rewritten as
 
 \begin{eqnarray}
 a^{\dagger}a + e^{\dagger}e + s^{\dagger}s + d^{\dagger}d = 1.\label{hdcondition}
 \end{eqnarray}
 It is easy to check that Eqn.(\ref{hdcondition}) is equivalent to 
 the following relations:  
 $ \{b_{\rm L}, b^{\dag}_{\rm L}\}=1$ and $\{b_{\rm R}, b^{\dag}_{\rm R}\} = 1$. 
 These two relations explicitly connect the extremely-strong 
 contact repulsion and Pauli exclusion principle 
 in the bosonic and fermionic systems, respectively \cite{girardeau}.      
 
 The particle number in one double well is given by
 
 \begin{eqnarray}
 b^{\dag}_{L}b_{L}+b^{\dag}_{R}b_{R}=2d^{\dag}d+s^{\dag}s+a^{\dag}a.
 \end{eqnarray}
 Since particles are hard-core bosons,    
 tunnelling of particles in $|1_{\rm L}1_{\rm R}\rangle$ state through the central barrier 
 is forbidden, and the population imbalance vanishes.
 If there is only one atom in a double well, the population imbalance reads

 \begin{eqnarray}
 \frac{1}{2}(b^{\dag}_{\rm L}b_{\rm L}-b^{\dag}_{\rm R}b_{\rm R})=\frac{1}{2}(s^{\dag}a+a^{\dag}s); \label{ordnt}
 \end{eqnarray}
 and the tunnelling term is rewritten as
 
 \begin{eqnarray}
 \frac{1}{2}(b^{\dag}_{\rm L}b_{\rm R}+{\rm h.c.})=\frac{1}{2}(s^{\dag}s-a^{\dag}a).\label{jospht}
 \end{eqnarray}
 If Bose-Einstein Condensate emerges, the Josephson tunnelling 
 may occur and the corresponding current reads
 
 \begin{eqnarray}
 -\frac{i}{2}(b^{\dag}_{\rm L}b_{\rm R}-b^{\dag}_{\rm R}b_{\rm L})=-\frac{i}{2}(a^{\dag}s-s^{\dag}a).
 \end{eqnarray}
 It is noted that the ordinary tunnelling in Eqn. (\ref{ordnt}) 
 and Josephson tunnelling in Eqn.(\ref{jospht}) are only 
 determined by singly occupied bases $|s\rangle$ and $|a\rangle$. 
 If the other two bases $|d\rangle$ and $|e\rangle$ have finite populations 
 induced by thermal fluctuations, 
 the tunnellings will decrease due to the constraint condition in Eqn. (\ref{hdcondition}).   

\section{Effective Hamiltonian and Self-Consistent Equations}

\subsection{Effective Hamiltonian}

\begin{figure}[t]
 \includegraphics[ scale=0.30 ]{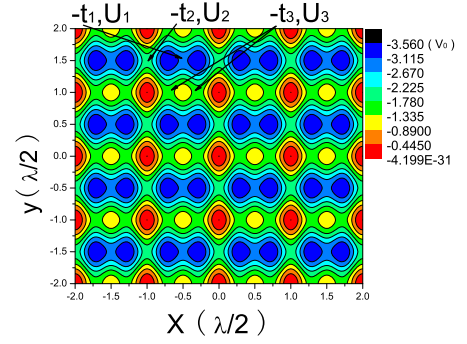}
 \caption{ (color online) Geometry of the double-well lattice.
 It is generated by superposition of 
 laser standing waves along $x$- and $y$-axis:
 $V(x,y)\!=\!-\!V_{0}{\rm sin}^{2}(k_{1}x)\!-\!V_{0}{\rm
 sin}^{2}(k_{2}x)\!-\!2V_{0}{\rm sin}^{2}(k_{3}y)$,
 where $2k_{1}\!\!\!=\!\!k_{2}\!\!\!=\!\!2k_{3}\!\!\!=\!\!2\pi/\lambda$ and $\lambda$
 is the laser wavelength.
 $t_{\sigma}$ are tunnelling amplitudes,
 and $U_{\sigma}$ ($\sigma\!=\!1,2,3$) are intra- ($\sigma\!=\!1$) and 
 inter-double-well ($\sigma\!=\!2,3$)  interactions.}\label{dwxy}
 \end{figure}

\begin{figure}[t]
 \includegraphics[ scale=0.45 ]{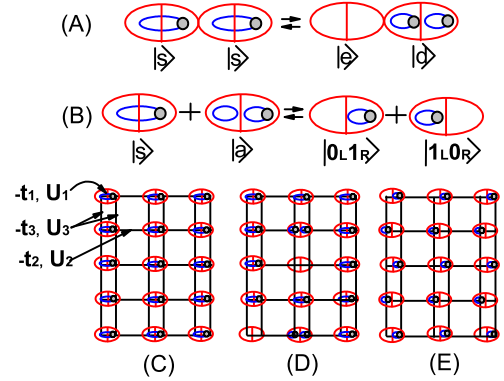}
 \caption{ (color online) Bosonic atoms in the double-well lattice 
 with filling factor one.
 (A) Simultaneous generation of a doubly occupied ($|d\rangle$) and empty ($|e\rangle$)
 state from the ground state.
 (B) and (E): Mixing between the symmetric ($|s\rangle$) and 
 anti-symmetric ($|a\rangle$) state
 results in two degenerate bases  
 $|0_{\rm L}1_{\rm R}\rangle$ and $|1_{\rm L}0_{\rm R}\rangle$.
 (C) Ground state composed of symmetric state ($|s\rangle$).
 (D) The double-well lattice in state that mixes $|d\rangle$, $|e\rangle$ and $|s\rangle$ .
 (E) The checkboard-like insulator state characterized by 
  wave vectors $(\pi,0)$ or $(0,\pi)$.}\label{dwhf}
 \end{figure}

 We now apply the hard-core form of the coupled representation to  
 bosonic atoms in the double-well lattice (see Fig. \ref{dwxy}).
 The double-well lattice is built up from  
 laser light standing waves along $x$- and $y$-axis:
 \begin{eqnarray}
   V(x,y)=-V_{0}{\rm sin}^{2}(k_{1}x)-V_{0}{\rm sin}^{2}(k_{2}x)-2V_{0}{\rm sin}^{2}(k_{3}y),\nonumber
 \end{eqnarray}
 where  
 $ 2k_{1}=k_{2}=2k_{3}=2\pi/\lambda $ with $\lambda$ 
 the wavelength of lasers.
 The second quantized Hamiltonian we considered here reads as
 
 \begin{eqnarray}
 H_{\rm DW} = 
  \sum_{r} H_{1,r} + H_{2,r} + H_{3,r} + H_{4,r},\label{model}
 \end{eqnarray}
 where 
 \begin{eqnarray}
 && H_{1,r} = 
   \frac{1}{2} U_{\rm L} n_{r,\rm L} (n_{r,\rm L}-1)        \nonumber\\
 &&\hspace{55pt} + ~~ \frac{1}{2} U_{\rm R} n_{r,\rm R} (n_{r,\rm R}-1);                      \nonumber\\
 %%%%%%%%%%%%%%%%
 && H_{2,r} =
   U_{1} n_{r,\rm L} n_{r,\rm R} 
 + U_{2} n_{r-\hat{x},\rm R} n_{r,\rm L}                         \nonumber\\
 &&\hspace{80pt} + ~~ U_{3} \sum_{\sigma} n_{r+\hat{y},\sigma} n_{r,\sigma};          \nonumber\\
 %%%%%%%%%%%%%%%%%
 && H_{3,r} = 
 -\mu_{r,\rm L} n_{r,\rm L}-\mu_{r,\rm R} n_{r,\rm R}.                        \nonumber\\
 %%%%%%%%%%%%%%%%
 && H_{4,r} = 
 - t_{1} b^{\dagger}_{r,\rm L} b_{r,\rm R}
 - t_{2} b^{\dagger}_{r-\hat{x},\rm R} b_{r,\rm L}  + {\rm h.c.}             \nonumber\\
 &&\hspace{85pt} - ~~ t_{3} \sum_{\sigma} b^{\dagger}_{r-\hat{y},\sigma}b_{r,\sigma}
 + {\rm h.c.};                                                               \nonumber
 %%%%%%%%%%%%%%
 \end{eqnarray}
 with $\sigma=\rm L,R$ and $b^{\dag}_{\rm L/R}$ ($b_{\rm L/R}$) 
 the creation (annihilation) operators of
 bosonic particles trapped in  the left  ($\rm L$) or
 the right ($\rm R$) side and 
 $n_{\rm L/R}=b^{\dag}_{\rm L/R}b_{\rm L/R}$ 
 the corresponding particle numbers. 
 $U_{\rm L/R}$ represent effective inter-particle contact potential in either side. 
 $t_{1}$ and $U_{1}$  measure the tunnelling amplitude and
 inter-particle interaction between the left and right sides within a double well. 
 $t_{2,3}$ and $U_{2,3}$ are the corresponding coupling parameters between 
 nearest-neighbour double wells as shown in Fig. \ref{dwhf}.
 The site-dependent chemical potential is denoted by $\mu_{r,\rm L/R}$.

 For deep lattice potential, interactions 
 $U_{1,2,3}$ between adjacent potential wells are much smaller than the 
 on-site ones $U_{\rm L,R}$ \cite{jaksch,scarola}.
 When $t_{1,2,3}\ll U_{\rm L/R}$ and filling factors are smaller than or equal to one,
 the bases with wells occupied by two or more particles 
 contribute much less to the ground state and can be safely omitted. 
 In such cases, particles are effectively hard-core bosons, satisfying
 $(b_{\rm L})^2=(b_{\rm R})^2=0$. 
 When the central barrier of a double well 
 is much lower than the barriers separating different double wells,
 the model Hamiltonian is further simplified by assuming 
 $t_{1} \gg t_{2,3}$, $ U_{1}\gg U_{2,3}$.
 In this work, we further assume $U_{i}=-\lambda t_{i}$ ($i=1,2,3$) for brevity. 
 
%%%%%%%%%%%%%%%%%%%%%%%%%%%%%%%%%%%%%%%%%%%%
%%%%%%%%%%%-- Fig.1 --%%%%%%%%%
%%%%%%%%%%%%%%%%%%%%%%%%%%%%%%%%%%%%%%%%%%%
%%%%%%%%%%%%%%%%%%%%%%%%%%%%%%%%%%%%%%%%%%%%%%%%

%%%%%%%%%%%%%%%%%%%%%%%%%%%%%%%%%%%%%%%%%%%%
%%%%%%%%%%%-- Fig.2 --%%%%%%%%%
%%%%%%%%%%%%%%%%%%%%%%%%%%%%%%%%%%%%%%%%%%%
 %%%%%%%%%%%%%%%%%%%%%%%%%%%%%%%%%%%%%%

%%%%%%%%%%%%%%%%%%%%%%%%%%%%%%%%%%%%%%%%%%%%
%%%%%%%%%%%-- effective Hamiltonian --%%%%%%%%%
%%%%%%%%%%%%%%%%%%%%%%%%%%%%%%%%%%%%%%%%%%%

 Particle tunnelling between two sides of a double well makes the single
 mode approximation unreasonable. 
 At least two modes per site (double well) 
 are required to construct an appropriate low-energy effective Hamiltonian for particles 
 in the double-well lattice.
 In the specific case with filling factor $\rho=1$ (one atom per double well), 
 and vanishing coupling between different double wells, 
 the ground state $\prod_{k} |s_{k}\rangle$ can be constructed from sDW bases $|s_{k}\rangle$, 
 with $k$ the site (double well) index, 
 since the sDW level is lower than and separated from the 
 other three levels by finite gaps as shown in Fig. \ref{dwhf}(C).
 When tunnelling $t_{2,3}$ between adjacent double wells  
 are strong enough to excite pairs of particles  
 $d$ and $e$ from the ground state $\prod_{k} |s_{k}\rangle$,
 mobility via $d$ and $e$ appears and the system are eventually in a fluid phase.
 The eigen wavefunction can now be build up from double-well bases that 
 mainly mix the three primary bases $|s\rangle$, $|d\rangle$, $|e\rangle$ 
 as  
 $
 [ \cos\theta s_{k}^{\dag}
 +\sin\theta (  d_{k}^{\dag} + e^{i\eta} e_{k}^{\dag} ]
 |\phi\rangle
 $.
 In this mixed state, pseudo particles $e$ and $d$ play a role similar to 
 what the hole and electron do in electronic crystal materials as 
 schematically shown in Fig. \ref{dwhf}(A) and (D).   
 On the other hand, repulsive interactions between atoms trapped in adjacent double wells 
 favour insulator state with characterization wave vectors 
 indicated by $(\pi,0)$ or $(0,\pi)$ as shown in Fig. \ref{dwhf}(E).
 If the filling factor $\rho$ changes from zero (empty) to
 $2$ (two atoms per double well), there will appear  
 other exotic commensurate and incommensurate insulator phases 
 such as $1/4$ or $1/8$ depleted insulator phases.

  When the filling factor is about one, the symmetric basis $|s\rangle$ is a good
  starting point to construct an variational wave function at low temperatures. 
  Applying the mean-field approximation 
  $\langle s^{\dagger} \rangle= \langle s \rangle=\bar{s}$ to the
  model Hamiltonian (\ref{model}),
  and neglecting the site dependence
  of chemical potentials $\mu_{r}$ by writing $\mu_{r}=\mu$, 
  i.e., the external confining potential and other kinds of
  well depth fluctuations are not considered here,
  the effective Hamiltonian can be written as 
  
  \begin{eqnarray}
  H_{\rm eff}(\bar{s},\mu,\nu)&=&
  H_{\rm  DW}(s^{\dagger},s\!\!\rightarrow\!\!\bar{s};\mu_{r}\!\!\rightarrow\!\!\mu)\!+\!2N\mu\rho\nonumber\\
  &&\!\!-\nu\sum_{r}(a^{\dagger}_{r}a_{r}\!+\!e^{\dagger}_{r}e_{r}
  \!+\!\bar{s}^{2}\!+\!d^{\dagger}_{r}d_{r}\!-\!1),\nonumber
  \end{eqnarray}  
  where the Lagrangian multiplier $\nu$ is introduced to
  assures the constraint condition in Eqn. (\ref{hdcondition}) under the mean field approximation. 
  If only bilinear terms of pseudo particles $d$, $e$ and $a$ are retained,
  the effective Hamiltonian is readily diagonalized
  by Fourier-Bogliubov transformations as
  
 \begin{eqnarray}
 H = \sum_{k,\alpha } \omega^{ \alpha }_{k}
 ( \alpha^{\dagger}_{k}\alpha_{k} + \frac{1}{2} ) + E_{0},\label{dh}
 \end{eqnarray}
 where 
 $\alpha = \tilde{d},\tilde{e},\tilde{a}$ and 
 \begin{eqnarray}
 E_{0}&=&-\frac{1}{4}\lambda\bar{s}^4(t_{2} + 2t_{3})
 + \bar{s}^2[-t_{1}+\frac{1}{2}\lambda(t_{2}+2t_{3})-\mu-\nu]\nonumber\\
 &&+\mu(\frac{3}{2}+\rho)
 +\frac{5}{2}\nu+\frac{1}{2}(\lambda-1)t_{1},\nonumber
 \end{eqnarray}
 and
 \begin{eqnarray}
 \omega^{\tilde{d},\tilde{e}}_{k} &=& \sqrt{A^{2}_{k} - \bar{s}^{4}\eta^{2}_{k}} \pm
 B_{k},\nonumber\\
 \omega^{\tilde{a}}_{k} &=& \sqrt{C^{2}_{k} - \bar{s}^{4}\xi^{2}_{k}},\nonumber
 \end{eqnarray}
 with 
 \begin{eqnarray}
 A_{k} &=& -\frac{\lambda}{2}t_{1} - \mu - \nu + \bar{s}^{2}\eta_{k},   	\nonumber\\ 		    
 B_{k} &=& -\frac{\lambda}{2}t_{1} - \mu, 	\nonumber\\
 C_{k} &=& t_{1}-\mu-\nu-\lambda\bar{s}^{2}(t_{2}+\!2t_{3})+\bar{s}^{2}\xi_{k},	\nonumber\\
 \eta_{k} &=& -t_{2}(\lambda+{\rm cos}{k_{x}})-2t_{3}({\rm cos}{k_{y}}+\lambda),	\nonumber\\
 \xi_{k} &=& \lambda(t_{2}{\rm cos}{k_{x}}-2t_{3}{\rm cos}{k_{y}}).	\nonumber
 \end{eqnarray}

%%%%%%%%%%%%%%%%%%%%%%%%%%%%%%%%%%%%%%%%%%%%%%%%%%%%%%%%%%%%%%%%%
%%%%%%%%%%%%%%%---hard-core bosonic statistics--%%%%%%%%%%%%%%%%%
%%%%%%%%%%%%%%%%%%%%%%%%%%%%%%%%%%%%%%%%%%%%%%%%%%%%%%%%%%%%%%%%%
\subsection{Hard-core Statistics}

 In contrast to identical bosons that obey the ordinary
 commutation relation, hard-core atoms follow the
 hard-core bosonic statistics.
 Hard-core bosonic statistics has been demonstrated to be
 a powerful tool in studying thermodynamic properties of
 triplon excitations in spin-dimer systems \cite{troyer,ruegg05}. 
 Similarly, we apply the hard-core statistics to
 hard-core bosonic atoms in the double-well optical lattice. 
 Consider the state subspace $S_{N}^{M}$ that has $M$ pseudo-particle excitation 
 ($\tilde{e}$,$\tilde{d}$,$\tilde{a}$ in the present paper) 
 over the ground state ($\prod_{k}|s_{k}\rangle$) in a $N$-sites lattice.
 If such excitations are regarded as identical bosons,  dimensions of $S_{N}^{M}$ read as
 \begin{eqnarray}
 g_{\rm IB}(N,M) = \left( \begin{array}{c}
 3N+M-1
 \\
 M
 \end{array}\right).\nonumber
 \end{eqnarray}   
 However, the real dimensions of $S_{N}^{M}$ should be
 \begin{eqnarray}
 g(N,M) = \left( \begin{array}{c}
 N
 \\
 M
 \end{array}\right)\times 3^{M}.\nonumber
 \end{eqnarray}
 In the case when $N \gg M \gg1$, the ratio
 \begin{eqnarray}
 \frac{ g(N,M) 
 }{ g_{\rm IB}(N,M) }\approx \exp\left[- 
 \left(\frac{2M}{\sqrt{3N}}\right)^{2} \right].\nonumber
 \end{eqnarray}
 When the typical energy of thermal fluctuations 
 is much less than the least excitation gap,
 the number of pseudo particles $M\ll \sqrt{N}$, and 
 the real dimensions of $S_{N}^{M}$ are approximately equal to that of identical bosons.
 The equilibrium properties at finite temperatures can be 
 discussed based on identical bosons.
 At higher temperatures, when $M\gtrsim\sqrt{N}$, the  
 identical bosons is not suitable to describe the equilibrium properties of such systems.
 
 Applying Troyer {\it et al.}'s method \cite{troyer} to double-well lattices,  
 the distribution of pseudo particles can be derived from partition function   
 of distinguishable bosonic particles in $S_{N}^{M}$.
 The partition function reads as
 \begin{eqnarray}
 Z_{\rm B}(N,M) = \sum_{k,\alpha}  e^{  -\beta\omega^{ \alpha }_{k}  },\nonumber
 \end{eqnarray}
 where $\alpha$ sums over $\tilde{d},\tilde{e},\tilde{a}$ 
 and $\beta=1/k_{B}T$ with $k_{B}$ the Boltzmann constant.
 The partition function $Z(N)$ of hard-core pseudo-particles can be obtained by 
 rescaling the dimensions of subspace $S_{N}^{M}$ as
 
 \begin{eqnarray}
 Z(N) &=& \sum_{M=0}^{N}\frac{ g(N,M) }{ (3N)^{M} } Z_{\rm B}(N,M)\nonumber\\
 &=&\left[
 1+\sum_{\alpha}\frac{1}{N}\sum_{k}  e^{  -\beta\omega^{ \alpha }_{k}  }
 \right]^{N}.\nonumber
 \end{eqnarray}
 The number of hard-core pseudo-particles per site can be written as
 
 \begin{eqnarray}
 n^{\alpha}_{k}=e^{-\beta\omega^{\alpha}_{k}}\Bigg/
 \Big[ 1+\frac{1}{N}\sum_{k,\alpha}e^{-\beta\omega^{\alpha}_{k}}  
 \Big ].
 \end{eqnarray} 
 where $\alpha$ sums over $\tilde{d}$, $\tilde{e}$, and $\tilde{a}$.
 
 \subsection{Self-consistent Saddle-point Equations}
 
 Calculations of the partition function for Hamiltonian (\ref{dh}) give the free energy per
 double well as
 
 \begin{eqnarray}
 f=E_{0}/N-\frac{1}{N}\sum_{k,\alpha}\beta
 {\rm ln}\left[\frac{1}{2}{\rm csch}\left(
 	\frac{\beta\omega^{\alpha}_{k}}{2}  \right)
 	\right].
 \end{eqnarray}
 The saddle-point equations of $f$ with respect to $\bar{s}^{2}$, $\mu$, and $\nu$ 
 can be self-consistently solved. 
 The self-consistent equations are written as
 \begin{widetext}
 \begin{eqnarray}
    &&-\!\frac{1}{2}\lambda(\bar{s}^{2}\!-\!1)(t_{2}\!+\!2t_{3})\!-\!\mu-\!\nu\!-\!t_{1}
 \!+\!\frac{1}{N}\sum_{k}\Big[(n^{\alpha}_{k}\!+\!n^{\beta}_{k}\!+\!1)
 \frac{A_{k}\eta_{k}\!-\!\bar{s}^{2}\eta^{2}_{k}}{\sqrt{A^{2}_{k}\!-\!\bar{s}^{4}\eta^{2}_{k}}}
 \!+\!(n^{\gamma}_{k}\!+\!\frac{1}{2})\frac{-\!\lambda C_{k}(t_{2}\!+\!t_{2})\!+\!C_{k}\xi_{k}\!-\!\bar{s}^{2}\xi^{2}_{k}}
 {\sqrt{C^{2}_{k}\!-\!\bar{s}^{4}\xi^{2}_{k}}} \Big]\!=\!0,\nonumber\\
    &&\frac{5}{2}\!-\!\bar{s}^2\!-\!\frac{1}{N}\sum_{k}\Big[(n^{\alpha}_{k}\!+\!n^{\beta}_{k}\!+\!1)
 \frac{A_{k}}{\sqrt{A^{2}_{k}\!-\!\bar{s}^{4}\eta^{2}_{k}}}
 \!+\!(n^{\gamma}_{k}\!+\!\frac{1}{2})\frac{C_{k}}
 {\sqrt{C^{2}_{k}\!-\!\bar{s}^{4}\xi^{2}_{k}}}\Big]\!=\!0,\nonumber\\
    &&-1\!+\!\rho\!+\!\frac{1}{N}\sum_{k}(n^{\beta}_{k}\!-\!n^{\alpha}_{k})\!=\!0.
 \end{eqnarray}
 \end{widetext}

 \section{Excitation Spectra and Thermodynamics}

%%%%%%%%%%%%%%%%%%%%%%%%%%%%%%%%%%%%%%%%%%%%%%%%%%%%%%%%%%%%%%%%%
%%%%%%%%%%%%%%%------excitation spectra---------%%%%%%%%%%%%%%%%%
%%%%%%%%%%%%%%%%%%%%%%%%%%%%%%%%%%%%%%%%%%%%%%%%%%%%%%%%%%%%%%%%%

 \subsection{Excitation Spectra}

\begin{figure}
 \includegraphics[ scale=0.75 ]{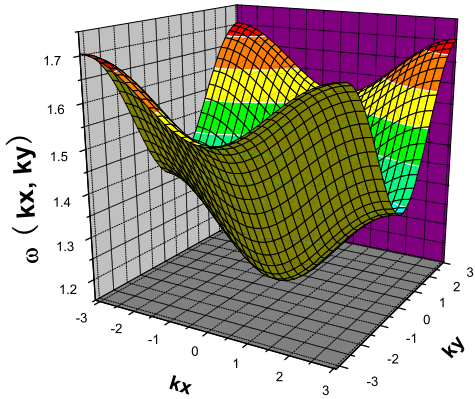}
 \caption{ (color online)  The excitation spectra of pseudo particles $\tilde{d}$  
 at $\lambda=-0.4$, $t_2/t_1=t_3/t_1=0.1$, and $\rho=1$.}\label{omgak}
 \end{figure}

\begin{figure}
 \includegraphics[ scale=0.75 ]{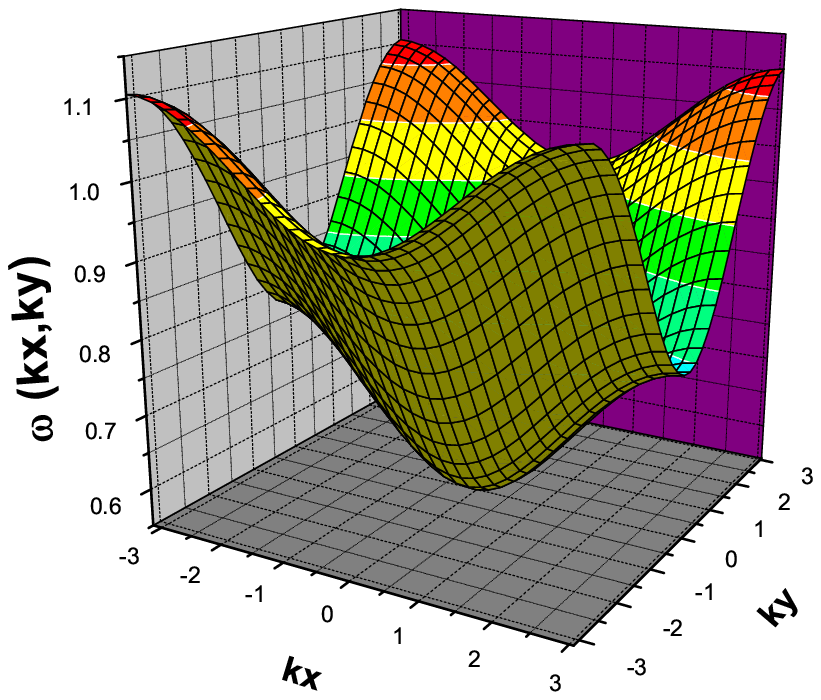}
 \caption{ (color online) The excitation spectra of pseudo particles $\tilde{e}$ 
 at $\lambda=-0.4$, $t_2/t_1=t_3/t_1=0.1$, and $\rho=1$.}\label{omgbk}
 \end{figure}

\begin{figure}
 \includegraphics[ scale=0.75 ]{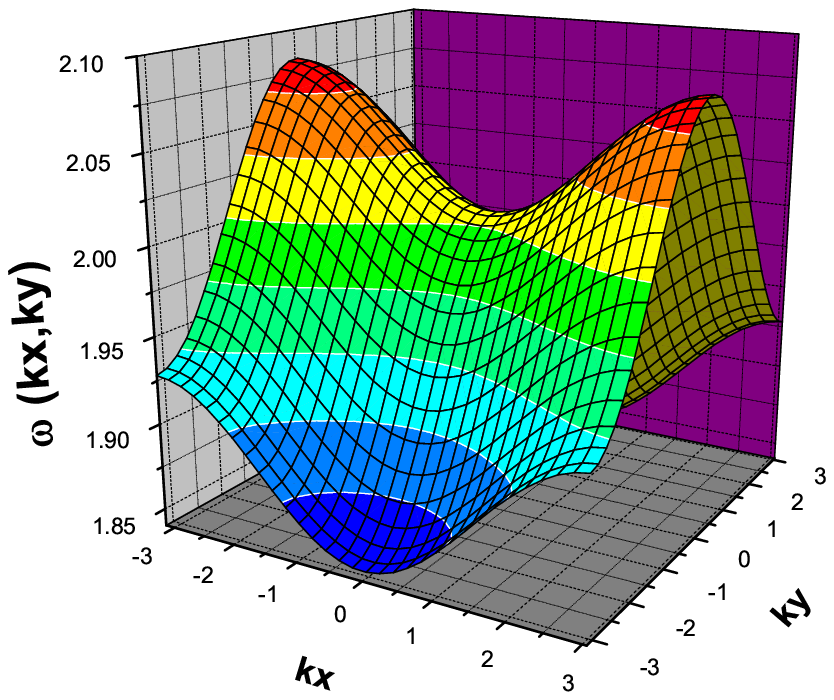}
 \caption{ (color online) The excitation spectra of pseudo particles $\tilde{a}$ 
 at $\lambda=-0.4$, with $t_2/t_1=t_3/t_1=0.1$, and $\rho=1$.}\label{omgck}
 \end{figure}

 Figs. \ref{omgak}-\ref{omgck} show the excitation spectra 
 at vanishing temperatures of hard-core bosonic system in the double-well lattice with
 $\lambda=-0.4$, $t_2/t_1=t_3/t_1=0.1$, and $\rho=1$. 
 It can be seen from theses figures that $\tilde{a}$ level is higher than 
 $\tilde{d}$ level and $\tilde{d}$ level is higher than $\tilde{e}$ level. 
 Hence, the lowest excitation seems to be $\omega_{k}^{\tilde{e}}$.
 However, under the condition of hard-core limit and 
 filling factor one, the excitation processes can 
 be expressed as $\bar{s}+\bar{s}\rightarrow \tilde{d}+\tilde{e}$,
 since the pseudo particles $\tilde{e}$ and $\tilde{d}$ are simultaneously excited. 
 Therefore, the lowest energy needed to create excitations 
 over the ground state is not the gap of $\tilde{e}$ 
 but the total energy required to generate 
 $\tilde{d}$ and $\tilde{e}$ pseudo particle pairs. 
 The average energy per particle reads as  
 $( \omega_{k}^{\tilde{e}}+\omega_{k}^{\tilde{d}} )/2
 =\sqrt{A^{2}_{k} - \bar{s}^{4}\eta^{2}_{k}}$.

\begin{figure}
 \includegraphics[ scale=0.38 ]{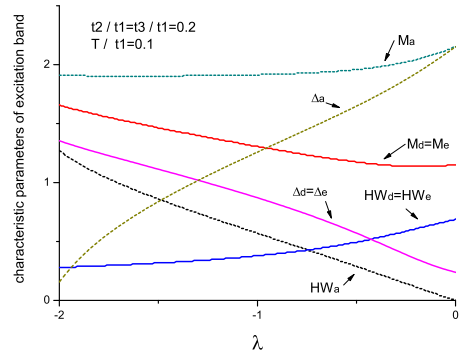}
 \caption{ (color online) Curves showing the dependence of gap $\Delta_{\alpha}$, 
 middle values $M_{\alpha}$, and half band widths $HW_{\alpha}$ of excitation spectra on
 the nearest-neighbour repulsive interactions
 with $t_2/t_1=t_3/t_1=0.2$, $T/t_1=0.1$ and $\rho=1$.}\label{gmh}
 \end{figure}

\begin{figure}
 \includegraphics[ scale=0.40 ]{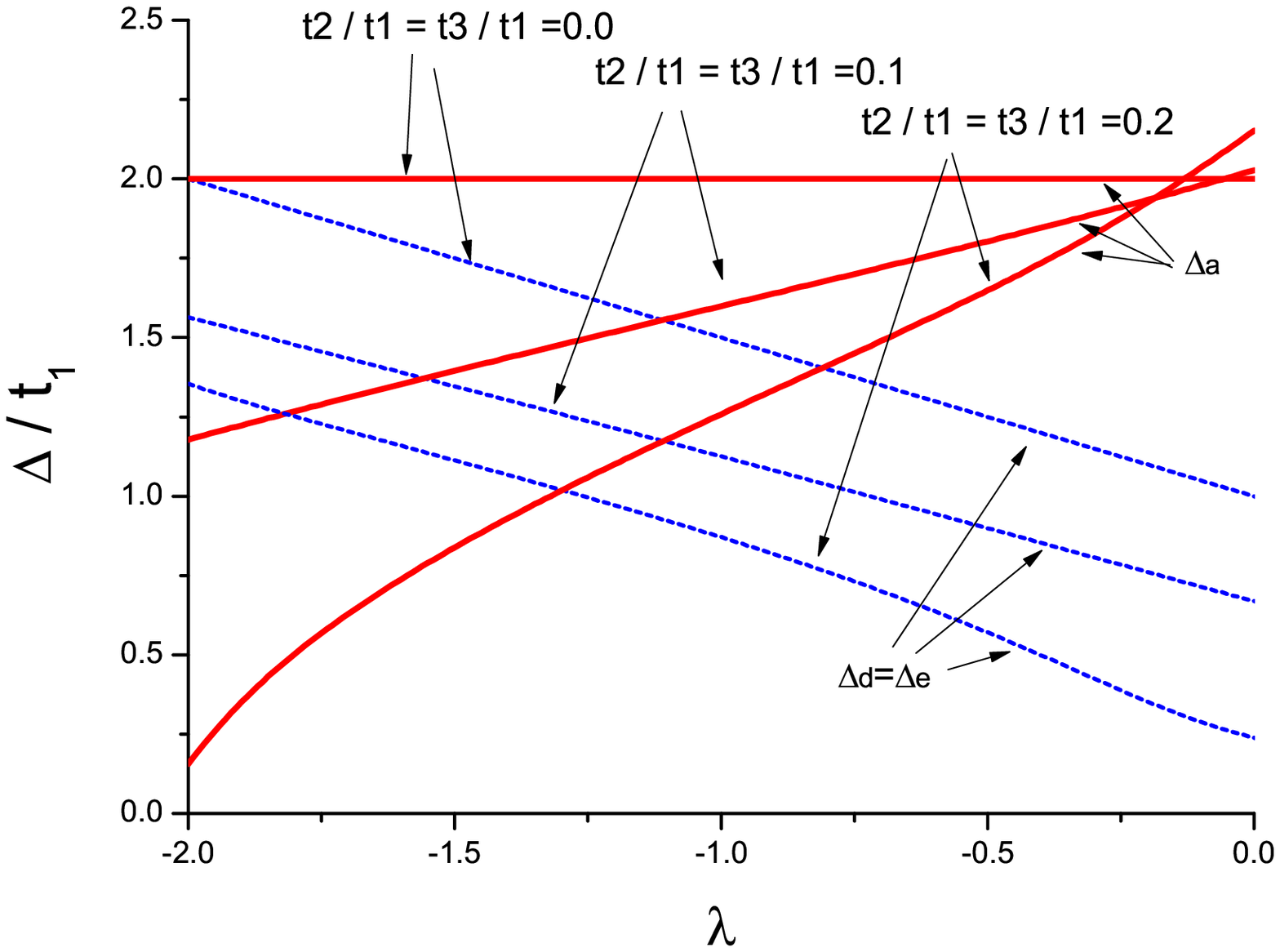}
 \caption{ (color online) Comparison of dependence of excitation gaps
 on the nearest-neighbour repulsive interactions
 with different tunnellings $t_2/t_1=t_3/t_1=0.0,0.1,0.2$,
 at $T/t_1=0.1$ and $\rho=1$.}\label{gl}
 \end{figure}

\begin{figure}[t]
 \includegraphics[ scale=0.40 ]{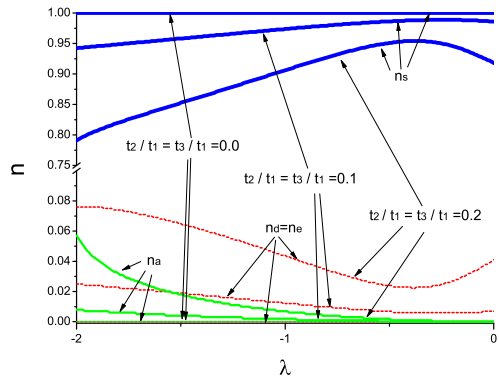}
 \caption{ (color online) Inter-particle interaction dependence of empty($n_{e}$), 
 singly ($n_{a}$, $n_{s}$) and doubly ($n_{d}$) occupied double-well numbers   
 at tunnellings $t_2/t1=t_3/t_1=0.0,0.1,0.2$,
 temperature $T/t_1=0.1$, and filling factor $\rho=1$.}\label{nlam}
 \end{figure}

 The $\lambda$ dependence of excitation gaps $\Delta_{\alpha}$, 
 middle values $M_{\alpha}$, and half band widths $HW_{\alpha}$ 
 is shown in Fig. \ref{gmh} and Fig. \ref{gl}. 
 Considering the degeneracy
 between $\omega^{\tilde{d}}_{k}$ and $\omega^{\tilde{e}}_{k}$, 
 the relationship $\mu=-\frac{1}{2}\lambda t_{1}$
 is satisfied. The gaps, middle values of
 the excitation spectra $\omega^{\tilde{d}}_{k}$ and $\omega^{\tilde{e}}_{k}$ are formulated by
 \begin{eqnarray}
 \Delta_{\tilde{d},\tilde{e}} &=& \sqrt{  |\nu| [|\nu|-2\bar{s}^{2}(\lambda+1)(t_{2}+2t_{3})]  }
 ~~(k_{x}=k_{y}=0),\nonumber\\
 M_{\tilde{d},\tilde{e}} &=& \sqrt{|\nu|[|\nu|-2\bar{s}^{2}\lambda(t_{2}+2t_{3})]}. \nonumber
 \end{eqnarray}
 As can be seen in Fig. \ref{gmh}, with $|\lambda|$ increasing from $0.5$,
 $M_{\tilde{d},\tilde{e}}$ is lifted while $HW_{\tilde{d},\tilde{e}}$ is narrowed.
 This is why the gap is lifted and the occupation number
 turns bigger (in contrast to $e^{ -\beta\omega^{\tilde{d},\tilde{e} } }$) (Fig. \ref{nlam})
 with increasing repulsions at fixed temperature.

 While strengthening the repulsive potential ($\lambda<0$), the gap $\Delta_{\tilde{a}}$
 of the antisymmetric singly occupied level decreases to zero, since the repulsions between nearest
 double wells keep the atoms away from each
 other. Consequently, the ground state of a two-fold degenerate checkboard-like
 insulator, which mixes the symmetric and anti-symmetric state
 $\prod_{r}({\rm cos}\theta s^{\dagger}_{r}+ {\rm sin}\theta a^{\dagger}_{r})
 |\phi\rangle$
 with $\theta\rightarrow\pm\pi/4$, can be expected.

 To explore the effect of $\nu$ on the excitation spectra, 
 $\lambda$ dependence of
 gaps with  $t_{2}=t_{3}=0.0,0.1,0.2$ is depicted in Fig. \ref{gl}. 
 When $t_2= t_3 =0$, the system is made up of isolated
 double wells and the eigen energy level of the antisymmetric 
 state does not depends on $\lambda$, as defined in Eqn. (\ref{dwbo}) and
 shown in Fig. \ref{gl}. Comparing with the the diagonalized
 excitation spectra $\omega^{\tilde{a}}_{k}$ in Eqn. (\ref{dh}),
 the relationship $|\nu|=(1-\frac{\lambda}{2})t_{1}$ can be reached.
 Substituting this expression into $\omega^{\tilde{d}}_{k}$ and $\omega^{\tilde{e}}_{k}$ ,
 we reobtain the linear relation
 $\Delta_{\tilde{d},\tilde{e}}=M_{\tilde{d},\tilde{e}}=|\nu|=(1-\frac{\lambda}{2})t_{1}$.

%%%%%%%%%%%%%%%%%%%%%%%%%%%%%%%%%%%%%%%%%%%%
%%%%%%%%%%%-- Fig.3 --%%%%%%%%%
%%%%%%%%%%%%%%%%%%%%%%%%%%%%%%%%%%%%%%%%%%%
%%%%%%%%%%%%%%%%%%%%%%%%%%%%%%%%%%%%%%%%%%%

%%%%%%%%%%%%%%%%%%%%%%%%%%%%%%%%%%%%%%%%%%%%
%%%%%%%%%%%-- Fig.4 --%%%%%%%%%
%%%%%%%%%%%%%%%%%%%%%%%%%%%%%%%%%%%%%%%%%%%
 %%%%%%%%%%%%%%%%%%%%%%%%%%%%%%%%%%%%%%%%%%%

%%%%%%%%%%%%%%%%%%%%%%%%%%%%%%%%%%%%%%%%%%%%
%%%%%%%%%%%-- Fig.5 --%%%%%%%%%
%%%%%%%%%%%%%%%%%%%%%%%%%%%%%%%%%%%%%%%%%%%
 %%%%%%%%%%%%%%%%%%%%%%%%%%%%%%%%%%%%%%%%%%%

%%%%%%%%%%%%%%%%%%%%%%%%%%%%%%%%%%%%%%%%%%%%
%%%%%%%%%%%-- Fig.6 --%%%%%%%%%
%%%%%%%%%%%%%%%%%%%%%%%%%%%%%%%%%%%%%%%%%%%
%%%%%%%%%%%%%%%%%%%%%%%%%%%%%%%%%%%%%%%%%%%

%%%%%%%%%%%%%%%%%%%%%%%%%%%%%%%%%%%%%%%%%%%%
%%%%%%%%%%%-- Fig.7 --%%%%%%%%%
%%%%%%%%%%%%%%%%%%%%%%%%%%%%%%%%%%%%%%%%%%%
 %%%%%%%%%%%%%%%%%%%%%%%%%%%%%%%%%%%%%%%%%%%

\subsection{Equilibrium Properties at Finite Temperatures}

\begin{figure}[t]
 \includegraphics[ scale=0.40 ]{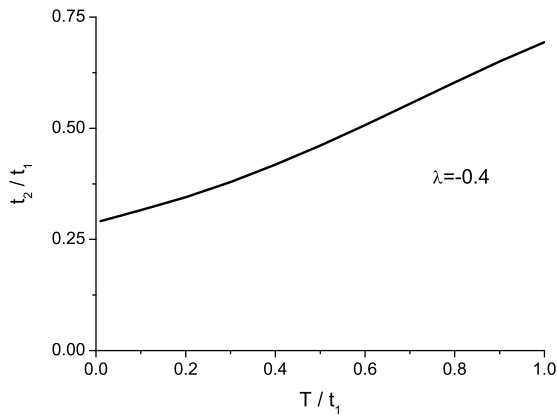}
 \caption{ Temperature (T) dependence of the critical 
 tunnellings between nearest double wells, $t_2$ and
 $t_{3}$, which collapse the lowest excitation gaps at $\lambda=-0.4$ and
 $\rho=1$.}\label{phase}
 \end{figure}

 Fig. \ref{phase} shows the temperature dependence of the critical 
 tunnellings $t_{2}$ and $t_{3}$, at $\lambda=-0.4$ and $\rho=1$. 
 The critical values are determined
 by setting the lowest excitation energy gap zero.
 In the region under the critical line shown in Fig. \ref{phase},
 the ground sate wave function is modified from 
 $\Psi_{1}=\prod_{i=1}^{N}|s\rangle$
 to 
 $\Psi'_{1}=[\sum_{r}( 1-\tilde{d}_{r}^{\dag}-\tilde{e}_{r}^{\dag}-\tilde{a}_{r}^{\dag} ) ]^{N} |\phi\rangle$.

\begin{figure}[t]
 \includegraphics[ scale=0.40 ]{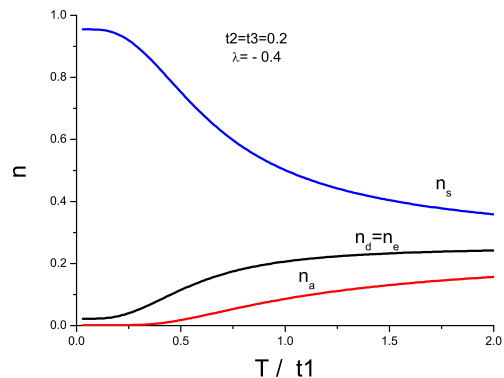}
 \caption{ (color online) Temperature dependence of empty($n_{e}$), 
 singly ($n_{a}$, $n_{s}$) and doubly ($n_{d}$) occupied double-well numbers   
 at tunnellings $t_2/t_1=t_3/t_1=0.2$,
 $\lambda=-0.4$, and filling factor $\rho=1$.}\label{nt}
 \end{figure}

 The critical tunnelling amplitude increases with temperature, which can be 
 explained as follows. Thermal fluctuations
 lead to finite occupations in excited levels $|\tilde{d}\rangle$ 
 and $|\tilde{{e}}\rangle$ (Fig. \ref{nt}), 
 and therefore suppress the influence of quantum tunnelling 
 so that a larger tunnelling amplitude 
 is required to collapse the excitation gap as is discussed 
 at the end of Section II. 
 It is noted that the number of pseudo particle $\tilde{d}$
 is the same to that of $\tilde{e}$, 
 suggesting that these two types of pseudo particles are simultaneously 
 created through the scattering process 
 $\bar{s}+\bar{s}\rightarrow \tilde{d}+\tilde{e}$.
 Even near vanishing temperatures at which 
 gapped excitations are few, a finite number of 
 doubly occupied and empty double wells still survive.
 That is due to the overlap between Wannier wave functions
 of nearest double wells characterized by 
 tunnellings $t_{\sigma}$ $(\sigma=2,3)$
 and the inter-atom interactions $-\lambda t_{\sigma}$.
 When temperature is high enough, the occupation number
 per Wannier level asymptotically equals $1/4$
 as shown in Fig. \ref{nt}, 
 which verifies the necessity of the hard-core statistics.

\begin{figure}[t]
 \includegraphics[ scale=0.40 ]{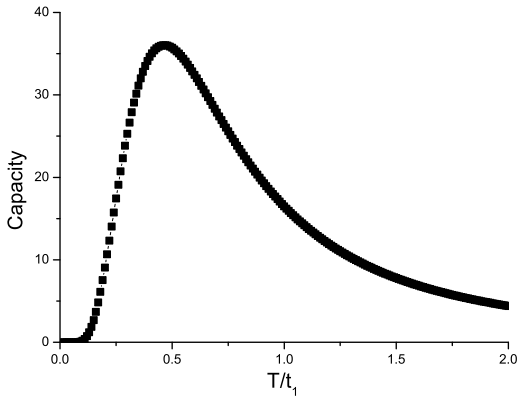}
 \caption{ Temperature dependence of heat capacity of hard-core bosonic atoms 
 in the double-well lattice
 at tunnellings $t_2/t_1=t_3/t_1=0.1$,
 $\lambda=-0.4$, and filling factor $\rho=1$.}\label{capct}
 \end{figure}

 Fig. \ref{capct} shows the temperature dependence of heat capacity.
 It is easy to see that a wide peak appears at temperature near $0.5t_{1}$.
 In the range of $T>0.5t_{1}$, 
 the heat capacity slowly decreases with temperature. 
 This reflects the hard-core nature of 
 bosonic atoms studied in this work. 
 While decreasing temperature from $0.5t_{1}$ to $0.125t_{1}$, 
 the capacity decreases rapidly since 
 the excitations gaps are finite at the given parameters.   
 When $T<0.125t_{1}$, the capacity is negligibly small 
 because all double wells are in $|s\rangle$ state except 
 contributions from few pseudo-particles $\tilde{d}$ and $\tilde{e}$.
 Although experimental results about the equilibrium properties mentioned above are 
 not available yet, there are quite a few experimental and theoretical works 
 on the thermodynamic properties of bosonic atoms in ordinary optical lattices \cite{garcia,gerbier}.
 In such systems, the influence of finite temperatures on the phase diagram was emphasized 
 by comparing numerical simulations with experimental results \cite{mahmud}.
 It can be expected that the effects of finite temperatures on the properties of atomic gases in 
 optical lattice will attract more attentions.

 %%%%%%%%%%%%%%%%%%%%%%%%%%%%%%%%%%%%%%%%%%%%
 %%%%%%%%%%%-- Fig.8 --%%%%%%%%%
 %%%%%%%%%%%%%%%%%%%%%%%%%%%%%%%%%%%%%%%%%%%

 %%%%%%%%%%%%%%%%%%%%%%%%%%%%%%%%%%%%%%%%%%%%
 %%%%%%%%%%%-- Fig.9 --%%%%%%%%%
 %%%%%%%%%%%%%%%%%%%%%%%%%%%%%%%%%%%%%%%%%%%
 
 %%%%%%%%%%%%%%%%%%%%%%%%%%%%%%%%%%%%%%%%%%%%
 %%%%%%%%%%%-- Fig.10 --%%%%%%%%%
 %%%%%%%%%%%%%%%%%%%%%%%%%%%%%%%%%%%%%%%%%%%

 %%%%%%%%%%%%%%%%%%%%%%%%%%%%%%%%%%%%%%%%%%%%
 %%%%%%%%%%%-- Fig.11 --%%%%%%%%%
 %%%%%%%%%%%%%%%%%%%%%%%%%%%%%%%%%%%%%%%%%%%
 
%%%%%%%%%%%%%%%%%%%%%%%%%%%%%%%%%%%%%%%%%%%%%%%%%%%%%
 
%%%%%%%%%%%%%%%%%%%%%%%%%%%%%%%%%%%%%%%%%%%%%%%%%%%%%

\subsection{Filling Factors Unequal to One}

To study excitation spectra and thermodynamic properties of 
other phases appearing at filling factors unequal to one, 
it is necessary to mix the bases defined in section II. 
SU(4) rotation transformations can be employed to accomplish such mixing, 
the same as  what is done in spin-dimer systems \cite{becker2001,vojta2013,penc2011}. 
The ground sate $\tilde{\Psi}$ can be obtained by minimizing the system energy  
$\langle\tilde{\Psi}|H|\tilde{\Psi}\rangle$, 
where $\tilde{\Psi}$ is the system wavefunction that is made up of 
double-well mixing bases determined by matrix elements of SU(4) transformations.
When the transformation matrix for the lowest double-well basis is determined, 
the rest double-well bases corresponding to excited levels 
can  also be obtained, as they are determined by the same set of variational parameters.
Based on the mixing bases constructed, other phases beyond the fluid 
and checkbaord-like insulator phases studied in this work 
could be systematically investigated, too.   
As has been stressed \cite{becker2001}, the orthogonality of mixed bases must be guaranteed in order 
to correctly describe other ordered phases, such as $1/4$ or $1/8$ depleted 
commensurate or incommensurate insulator phases. 
Such generalized cases are quite interesting and will be regarded 
as our next work. 

%%%%%%%%%%%%%%%%%%%%%%%%%%%%%%%%%%%%%%%%%%%%%%%%%%%%%%%%%%%%%

 \section{Conclusions}
 
 In summary, a generalized coupled representation for bosonic atoms 
 in double-well lattices has been obtained by exploiting 
 the mapping relationship between atomic occupation state 
 and coupled bases for spin-dimer systems.
 Then the coupled representation is applied to the hard-core case
 with filling factor one. The 
 excitation spectra and thermodynamic properties of such systems are investigated.
 Starting with a variational ground state wavefunction made of singly 
 occupied symmetric double-well bases, the excitation processes
 are described by creating pseudo-particles pairs $\tilde{d}$ and $\tilde{e}$, 
 and pseudo-particles $\tilde{a}$. 
 $\tilde{d}$, $\tilde{e}$ and $\tilde{a}$
 correspond to doubly occupied and empty double wells 
 and antisymmetric singly occupied double wells, respectively.
 It is demonstrated that hard-core statistics is required to 
 precisely describe the equilibrium properties of bosonic atoms 
 in double-well lattices at finite temperatures. 
 The critical tunnelling amplitudes increase monotonically with temperature.     
 This behaviour is qualitatively explained based on the effects of 
 thermodynamic fluctuations on the quantum tunnelling.
 The heat capacity and particle numbers based on the hard-core statics 
 are also calculated which need future experimental verifications.

\end{document}